\documentclass{JHEP3}
\preprint{}

\skip\footins = 1\bigskipamount plus 2pt minus 4pt

\usepackage{euscript,bbold}

\def\c{\chi}
\def\cL{{\cal{L}}}
\renewcommand\Re{{\cal R}}
\newcommand\U{\mathop{\rm {}U}\nolimits}
\newcommand\SU{\mathop{\rm SU}\nolimits}
\newcommand\SO{\mathop{\rm SO}\nolimits}

\title{General supersymmetric $AdS_5$ black holes}

\author{Jan B. Gutowski\\
Mathematical Institute\\
Oxford, OX1 3LB, U.K.\\
E-mail: \email{gutowski@maths.ox.ac.uk}}

\author{Harvey S. Reall\\
Kavli Institute for Theoretical Physics, University of California\\
Santa Barbara, CA 93106-4030, U.S.A.\\
E-mail: \email{reall@kitp.ucsb.edu}}

\abstract{Supersymmetric, asymptotically $AdS_5$, black hole solutions
  of five dimensional gauged supergravity coupled to arbitrarily many
  abelian vector multiplets are presented. The general nature of
  supersymmetric solutions of this theory is discussed. All maximally
  supersymmetric solutions of this theory (with or without gauging)
  are obtained.}

\begin{document}

\section{Introduction}\label{section1}

An obvious requirement of any theory of quantum gravity is to provide
a microscopic explanation of black hole entropy. To some extent, the
AdS/CFT correspondence~\cite{maldacena:98} has done this: the
microstates of an asymptotically anti-de Sitter (AdS) black hole are
states of a dual CFT. In three dimensions, it is possible to calculate
the entropy of black holes by counting CFT states with the same
quantum numbers as the black hole using the Cardy
formula~\cite{strominger:98}. However, an analagous calculation in
more than three dimensions has been hindered by our poor understanding
of strongly coupled quantum field theory.

A potential way of overcoming this obstacle is to consider
asymptotically AdS black holes preserving enough supersymmetry that
the corresponding CFT states belong to short superconformal
multiplets. The number of such states with given quantum numbers is
not expected to change as the CFT coupling is varied, so by counting
them at weak coupling it should be possible to reproduce the black
hole entropy.

Four dimensional, supersymmetric, asymptotically AdS, black holes were
presented in~\cite{kostalecky:96}. However, a CFT calculation of their
entropy does not seem possible because so little is known about the
dual CFT in this case. Seven dimensional supersymmetric AdS black
holes were constructed in~\cite{gauntlett:01}. Once again, our
ignorance of the dual CFT gives us little hope of being able to
calculate their entropy.

This leads us to $D=5$, where our understanding of the CFT, ${\cal
  N}=4$ $\SU(N)$ super Yang-Mills theory, is much better.  The first
examples of supersymmetric, asymptotically $AdS_5$, black holes were
recently obtained in~\cite{gutowski:04} as solutions of minimal $D=5$
gauged supergravity. These solutions have to rotate,\pagebreak[3] just like
supersymmetric AdS black holes in $D=3,4$ (but unlike $D=7$). They
preserve at least two supersymmetries, and their mass $M$ and angular
momenta $J_{1,2}$ are functions of their electric charge $Q$:
$M=M(Q)$, $J_1 = J_2 = J(Q)$. These black holes should correspond to
$\SU(N)$ singlet CFT states on $R \times S^3$. The CFT state/operator
correspondence maps such states to gauge-invariant local operators at
the origin of (euclidean) $R^4$~\cite{witten:98}.  In order to
identify the local operators corresponding to the black hole, the
solution must first be oxidized to give an asymptotically $AdS_5
\times S^5$ solution of type-IIB supergravity. This can be done using
results of~\cite{chamblin:99}. One finds that the dual CFT operators
are $1/16$ BPS,  have $\SO(4) = \SU(2)_L \times \SU(2)_R$ spins $J_L =
J(Q)$, $J_R=0$ and R-charge given by the $\SU(4)$ weight vector
$(Q,Q,Q)$ (for suitably normalized $Q$). We expect that it should be
possible to reproduce the black hole entropy by counting such
operators at weak coupling.

It is natural to expect that there should be more general
supersymmetric black hole solutions for which the corresponding CFT
operators have R-charge $(Q_1,Q_2,Q_3)$. To see how these should
arise, note that type-IIB supergravity can be dimensionally reduced on
$S^5$ to give ${\cal N}=1$ $D=5$ gauged supergravity coupled to three
abelian vector multiplets, with gauge group
$\U(1)^3$~\cite{cvetic:99}.\footnote{Strictly speaking, there are only
two vector multiplets since the diagonal $\U(1)$ arises from the
graviphoton.}  This theory has eight supercharges.  Black holes
carrying $\U(1)$ charges $Q_i$ will correspond to CFT operators with
R-charge $(Q_1,Q_2,Q_3)$. The goal of the present paper is to find
supersymmetric, asymptotically $AdS_5$, black hole solutions with
these charges.

Although we are mainly interested in the $\U(1)^3$ theory, we shall
work within the more general framework of ${\cal N}=1$ $D=5$ gauged
supergravity coupled to arbitrarily many abelian vector
multiplets~\cite{gunaydin:85}. One possible way to find black hole
solutions would be to use the results of~\cite{gutowski:04} to
motivate an Ansatz for the fields of this theory, and then examine the
circumstances under which this Ansatz admits a super-covariantly
constant spinor. However, we shall adopt a more systematic approach in
which we analyze the general nature of supersymmetric solutions of
this theory. This approach was first used in~\cite{tod:83} for minimal
${\cal N}=2$, $D=4$ supergravity and has been applied recently to
minimal $D=5$~\cite{gauntlett:02} and $D=6$~\cite{gutowski:03}
supergravity, minimal $D=5$ gauged supergravity~\cite{gauntlett:03},
and minimal ${\cal N}=2$, $D=4$ gauged
supergravity~\cite{caldarelli:03}.  It has also been used in $D=11$
supergravity~\cite{gauntlett:03a, gauntlett:03b}.  In this way, not
only do we find the black hole solutions, we also obtain a general
framework which may be used to construct interesting new solutions,
many of which cannot be found easily by guessing Ans\"atze.

In section~\ref{section2} of this paper, we generalize some of the
results of~\cite{gauntlett:02, gauntlett:03} to include arbitrarily
many abelian vector multiplets.  Our first result is a complete
classification of all maximally supersymmetric solutions in both the
gauged and ungauged theory. In the gauged theory, the only maximally
supersymmetric solution is $AdS_5$ with vanishing gauge fields and
prescribed values for the scalars. In the ungauged theory, we find
that the scalars take arbitrary constant values but once these values
have been chosen, the maximally supersymmetric solutions are in
one-to-one correspondence with the maximally supersymmetric solutions
of the minimal theory, which were given in~\cite{gauntlett:02}.

We then turn our attention to general supersymmetric solutions. Just
as in the minimal theories~\cite{gauntlett:02, gauntlett:03}, such
solutions possess a globally defined Killing vector field that is
either everywhere null, or timelike somewhere. We consider only
solutions in the latter class, and show that they admit a
hyper-K\"ahler or K\"ahler structure in the ungauged and gauged
theories respectively, again just as in the minimal
theories~\cite{gauntlett:02, gauntlett:03}. We determine a general
form for such solutions of the gauged theory and show that they must
preserve at least two supersymmetries.

In section~\ref{section3}, we use these results to derive
supersymmetric black hole solutions of ${\cal N}=1$, $D=5$ gauged
supergravity coupled to arbitrarily many abelian vector
multiplets. Just like the solutions of~\cite{gutowski:04}, these are
asymptotically AdS solutions that are parametrized by their electric
charges, and have two equal angular momenta. We pay special attention
to solutions of the $\U(1)^3$ theory mentioned above since these can
be oxidized to give BPS, asymptotically $AdS_5 \times S^5$ solutions
of IIB supergravity.

We should note that supersymmetric solutions of $D=5$ gauged
supergravity coupled to vector multiplets have been discussed
before. In particular, there have been attempts at finding
supersymmetric black hole solutions of this theory, but these attempts
only produced solutions with naked singularities~\cite{behrndt:98} or
naked closed timelike curves~\cite{klemm:01a}. Some black string
solutions were presented in~\cite{convent:03}.

The reader interested only in the black hole solutions (and not their
derivation) should jump to subsection~\ref{subsec:properties}.

\section{Supersymmetric solutions of ${\cal N}=1$ supergravity}
\label{section2}

\subsection{${\cal N} = 1$ supergravity}\label{section2.1}

The action of ${\cal N}=1$ $D=5$ gauged supergravity coupled to $n$
abelian vector multiplets with scalars taking values in a symmetric
space is~\cite{gunaydin:85}\footnote{We use a negative signature
metric.}
\begin{equation}
 S = {1 \over 16 \pi G} \int \left( -{}^5 R + 2 \chi^2 {\cal{V}} -Q_{IJ} F^I
 \wedge *F^J +Q_{IJ} dX^I \wedge
 \star dX^J -{1 \over 6} C_{IJK} F^I \wedge F^J \wedge A^K \right)
\end{equation}
where $I,J,K$ take values $1, \ldots ,n$ and $F^I=dA^I$.  $C_{IJK}$
are constants that are symmetric on $IJK$ and obey
\begin{equation}
\label{eqn:jordan}
C_{IJK} C_{J' (LM}
C_{PQ) K'} \delta^{J J'} \delta^{K K'} = {4 \over 3} \delta_{I (L}
C_{MPQ)}\,.
\end{equation}
See~\cite{convent:03} for a more detailed recent discussion of this theory.
The $X^I$ are scalars which are constrained
via
\begin{equation}
\label{eqn:conda}
{1 \over 6}C_{IJK} X^I X^J X^K=1\,.
\end{equation}
We may regard the $X^I$ as being functions of $n-1$ unconstrained
scalars $\phi^a$. It is convenient to define
\begin{equation}
X_I \equiv {1 \over 6}C_{IJK} X^J X^K
\end{equation}
so that the condition~({\ref{eqn:conda}}) becomes
\begin{equation}
X_I X^I =1\,.
\end{equation}
In addition, the coupling $Q_{IJ}$ depends on the scalars via
\begin{equation}
Q_{IJ} = {9 \over 2} X_I X_J -{1 \over 2}C_{IJK} X^K
\end{equation}
so in particular
\begin{equation}
Q_{IJ} X^J = {3 \over 2} X_I\,,
\qquad
Q_{IJ} \partial_a X^J = -{3 \over 2} \partial_a X_I\,.
\end{equation}
The constraints~({\ref{eqn:jordan}}) are sufficient to ensure that the
matrix $Q_{IJ}$ is invertible with inverse $Q^{IJ}$ given by
\begin{equation}
Q^{IJ} = 2 X^I X^J -6 C^{IJK} X_K
\end{equation}
where
\begin{equation}
C^{IJK} \equiv \delta^{II'} \delta^{JJ'} \delta^{KK'} C_{I'J'K'}\,.
\end{equation}
It is then straightforward to show that
\begin{equation}
X^I = {9 \over 2} C^{IJK} X_J X_K\,.
\end{equation}
The scalar potential can be written as
\begin{equation}
{\cal{V}} = 27 C^{IJK} V_I V_J X_K
\end{equation}
where $V_I$ are constants.

For a bosonic background to be supersymmetric there must be a
spinor\footnote{We use symplectic Majorana spinors. Our conventions
  are the same as~\cite{gauntlett:02}.}  $\epsilon^a$ for which the
supersymmetry variations of the gravitino and dilatino vanish. For the
gravitino this requires
\begin{equation}
\label{eqn:grav}
\left[\nabla_\mu +{1 \over 8}X_I(\gamma_\mu{}^{\nu \rho}
-4 \delta_\mu{}^\nu \gamma^\rho) F^I{}_{\nu \rho} \right]
\epsilon^a-{\chi  \over 2} V_I (X^I \gamma_\mu-3 A^I{}_\mu) \epsilon^{ab}
\epsilon^b=0
\end{equation}
and for the dilatino it requires
\begin{equation}
\label{eqn:dil}
\left[ \left( {1 \over 4}Q_{IJ} \gamma^{\mu \nu} F^J{}_{\mu \nu}
+{3  \over 4} \gamma^\mu \nabla_\mu X_I \right) \epsilon^a
-{3 \chi \over 2} V_I  \epsilon^{ab} \epsilon^b \right]  {\partial X^I \over
\partial \phi^a} = 0\,.
\end{equation}
The Einstein equation is
\begin{equation}
\label{eqn:ein}
{}^5 R_{\alpha \beta} +Q_{IJ} F^I{}_{\alpha \lambda} F^J{}_\beta{}^\lambda
-Q_{IJ} \nabla_\alpha X^I \nabla_\beta X^J-{1 \over 6}g_{\alpha \beta}
\left(4 \chi^2 V +Q_{IJ} F^I{}_{\mu \nu} F^{J \mu \nu} \right) =0\,.
\end{equation}
The Maxwell equations (varying $A^I$) are
\begin{equation}
\label{eqn:gauge}
d \left(Q_{IJ} \star F^J \right)=-{1 \over 4}C_{IJK} F^J \wedge F^K\,.
\end{equation}
The scalar equations (varying $\phi^a$) are
\begin{eqnarray}
\label{eqn:scal}
\bigg[{-}d ( \star dX_I )+ \bigg( X_M X^P C_{NPI}-{1 \over 6}
C_{MNI} \bigg) (F^M \wedge  \star F^N -dX^M \wedge \star dX^N ) -
\hphantom{\bigg] {\partial X^I \over \partial \phi^a}}\! &&
\\
-6 \chi^2 C^{MPQ} V_M V_P C_{QIJ}X^J {\rm dvol} \bigg]
{\partial X^I \over \partial \phi^a} &=& 0\,.\qquad
\nonumber
\end{eqnarray}
If a quantity $L_I$ satisfies $L_I \partial_a X^I = 0$ then there must
be a function $M$ such that $L_I = M X_I$. This implies that the
dilatino equation~({\ref{eqn:dil}}) can be simplified to
\begin{equation}
\label{eqn:newdil}
\left[ \left({1 \over 4}Q_{IJ}-{3 \over 8}X_I X_J \right)
F^J{}_{\mu \nu} \gamma^{\mu \nu}
+{3 \over 4} \gamma^\mu \nabla_\mu X_I \right] \epsilon^a
+{3 \chi \over 2}\left(X_I V_J X^J -V_I \right) \epsilon^{ab} \epsilon^b =0\,,
\end{equation}
and the scalar equation can be written as
\begin{eqnarray}
-d \left(\star dX_I \right) + \left({1 \over 6} C_{MNI} -{1 \over
2}X_I C_{MNJ} X^J \right) dX^M \wedge \star dX^N + &&
\nonumber\\
+ \left( X_M X^P C_{NPI}-{1 \over 6}C_{MNI}-6 X_I X_M X_N+{1 \over 6}
X_I C_{MNJ} X^J \right) F^M \wedge \star F^N + &&
\nonumber\\
+ 6 \chi^2 \left( -C^{MPQ}V_M V_P C_{QIJ} X^J +6 X_I C^{MPQ} V_M V_P
X_Q \right) {\rm dvol} &=& 0\,.\qquad
\end{eqnarray}

\subsection{Maximal supersymmetry}\label{section2.2}

All maximally supersymmetric solutions of the minimal ungauged and
gauged $D=5$ supergravity theories were explicitly obtained
in~\cite{gauntlett:02} and~\cite{gauntlett:03} respectively. In the
gauged theory, the unique maximally supersymmetric solution is $AdS_5$
with vanishing gauge field. The ungauged theory has a more complicated
set of maximally supersymmetric solutions.\footnote{It was conjectured
  in~\cite{gauntlett:02} that some of the maximally supersymmetric
  solutions obtained there are isometric. This conjecture was proved
  in~\cite{fiol:03}.}

We shall now identify all maximally supersymmetric solutions of ${\cal
  N}=1$ $D=5$ supergravity coupled to $n$ vector multiplets. First we
examine the dilatino equation~({\ref{eqn:newdil}}). For maximal
supersymmetry, this equation must impose no algebraic constraints on
the Killing spinor, which implies that the terms with two, one and
zero gamma matrices must vanish independently. This gives
\begin{equation}
dX_I =0
\end{equation}
so the scalars $X_I$ (and hence also $X^I$) are constant:
\begin{equation}
 X_I \equiv \bar{X}_I.
\end{equation}
Moreover,
\begin{equation}
\left({1 \over 4}Q_{IJ}-{3 \over 8}\bar{X}_I \bar{X}_J \right)F^J  = 0\,,
\end{equation}
which implies that (at least locally)
\begin{equation}
A^I = \bar{X}^I A
\end{equation}
for some 1-form $A$. Lastly, in the gauged theory ($\chi \ne 0$) we
also obtain
\begin{equation}
 \bar{X}_I = \xi^{-1} V_I\,,
\end{equation}
where
\begin{equation}
 \xi^3 = \frac{9}{2} C^{IJK} V_I V_J V_K\,.
\end{equation}
In the ungauged theory, the values of $\bar{X}_I$ are arbitrary.  It
is convenient to define
\begin{equation}
 F=dA\,.
\end{equation}
Upon substituting these identities back into the equations of motion
and gravitino equation, we observe that the scalar equation holds
automatically. The other equations simplify to
\begin{eqnarray}
{}^5 R_{\alpha \beta} +{3 \over 2} F_{\alpha \lambda}
F_\beta{}^\lambda -{1 \over 2} g_{\alpha \beta}
\left(8 \c^2 \xi^2 +{1 \over 2} F_{\lambda \mu} F^{\lambda \mu}\right) &=& 0\,,
\\
d \star F + F \wedge F &=& 0\,,
\\
\left( \nabla_\mu +{1 \over 8} (\gamma_\mu{}^{\nu \rho} -4
\delta_\mu{}^\nu \gamma^\rho)F_{\nu \rho}\right)
\epsilon^a -{\c\xi \over 2}(\gamma_\mu -3A_\mu)\epsilon^{ab}\epsilon^b &=&0\,.
\end{eqnarray}
However, these are merely the equations of motion and gravitino
equation of the \emph{minimal} supergravity.\footnote{To compare with
  the conventions of~\cite{gauntlett:02, gauntlett:03} one must make
  the replacements $A \rightarrow (2 / \sqrt{3})A$ and $\c \xi
  \rightarrow \c/ (2 \sqrt{3})$.}  Hence the maximally supersymmetric
solutions of the non-minimal theory are in one-to-one correspondence
with those of the minimal theory.\footnote{Strictly speaking, in the
  ungauged theory the solutions are also parametrized by the values of
  the constants $\bar{X}_I$.} In particular, in the gauged theory, the
only maximally supersymmetric solution is $AdS_5$ with radius $\ell$
given by
\begin{equation}
 \frac{1}{\ell} = \chi \xi\,.
\end{equation}

\subsection{General supersymmetric solutions}\label{section2.3}

Following~\cite{gauntlett:02}, our strategy for determining the
general nature of bosonic supersymmetric solutions is to analyse the
differential forms that can be constructed from a (commuting) Killing
spinor. We first investigate algebraic properties of these forms, and
then their differential properties.

From a single commuting spinor $\epsilon^a$ we can construct a scalar
$f$, a 1-form $V$ and three 2-forms $\Phi^{ab} \equiv \Phi^{(ab)}$:
\begin{equation}
 f \epsilon^{ab} = \bar{\epsilon}^a \epsilon^b\,,
\qquad
 V_\alpha \epsilon^{ab} = \bar{\epsilon}^a \gamma_\alpha \epsilon^b\,,
\qquad
 \Phi^{ab}_{\alpha \beta} = \bar{\epsilon}^a \gamma_{\alpha \beta}
 \epsilon^b\,.
\end{equation}
$f$ and $V$ are real, but $\Phi^{11}$ and $\Phi^{22}$ are complex
conjugate and $\Phi^{12}$ is imaginary. It is convenient to work with
three real two-forms $J^{(i)}$ defined by
\begin{equation}
 \Phi^{(11)} = J^{(1)} + i J^{(2)}\,,
\qquad
 \Phi^{(22)} = J^{(1)} - i J^{(2)}\,,
\qquad
 \Phi^{(12)} = - i J^{(3)}\,.
\end{equation}
It will be useful to record some of the algebraic identities that can
be obtained from the {}Fierz identity~\cite{gauntlett:02}:
\begin{eqnarray}
\label{eqn:Vsq}
 V_{\alpha} V^{\alpha} &=& f^2\,,
\\
\label{eqn:XwedgeX}
 J^{(i)} \wedge J^{(j)} &=& -2\delta_{ij} f \star V\,,
\\
\label{eqn:VdotX}
i_V J^{(i)} &=& 0\,,
\\
\label{eqn:VstarX}
i_V \star J^{(i)} &=& - f J^{(i)}\,,
\\
\label{eqn:XcontX}
J^{(i)}_{\gamma \alpha} J^{(j) \gamma}{}_{\beta} &=& \delta_{ij} \left(
 f^2 \eta_{\alpha\beta} - V_{\alpha} V_{\beta} \right) + \epsilon_{ijk} f
 J^{(k)}_{\alpha\beta}
\end{eqnarray}
where $\epsilon_{123} = +1$ and, for a vector $Y$ and $p$-form $A$,
$(i_Y A)_{\alpha_1, \ldots, \alpha_{p-1}} \equiv Y^{\beta} A_{\beta
  \alpha_1, \ldots, \alpha_{p-1}}$. {}Finally, we have
\begin{eqnarray}
\label{eqn:Vproj}
 V_{\alpha} \gamma^\alpha \epsilon^a &=& f \epsilon^a\,
\\
\label{eqn:Phiproj}
\Phi^{ab}_{\alpha \beta} \gamma^{\alpha \beta} \epsilon^c &=& 8f
\epsilon^{c(a} \epsilon^{b)} \,.
\end{eqnarray}
Equation~(\ref{eqn:Vsq}) implies that $V$ is timelike, null or zero. The
final possibility can be eliminated using arguments
in~\cite{gauntlett:02,reall:03}.

We now turn to the differential conditions that arise because
$\epsilon$ is a Killing spinor. We differentiate $f$, $V$, $\Phi$ in
turn and use~({\ref{eqn:grav}}).  Starting with $f$ we find
\begin{equation} \label{eqn:df}
 df = -i_V \left(X_I F^I \right),
\end{equation}
which implies $\cL_V f=0$ where $\cL$ denotes the Lie derivative.
Next, differentiating $V$ gives
\begin{equation}
 D_{(\alpha} V_{\beta)} = 0\,,
\end{equation}
so $V$ is a Killing vector, and
\begin{equation}
\label{eqn:dV}
 dV = 2f X_I F^I + X_I \star (F^I \wedge V) +2 \c V_I X^I J^{(1)}\,.
\end{equation}
Finally, differentiating $J^{(i)}$ gives
\begin{eqnarray}
\label{eqn:dPhi}
 D_\alpha J^{(i)}_{\beta\gamma} &=& -{1 \over 2} X_I \left[ 2
F^I{}_{\alpha}{}^{\delta} \left( \star J^{(i)} \right)_{\delta\beta\gamma} -2
F^I{}_{[\beta}{}^{\delta} \left( \star J^{(i)} \right)_{\gamma] \alpha
 \delta} + \eta_{\alpha [\beta} {}F^{I \delta \epsilon} \left( \star J^{(i)}
 \right)_{\gamma] \delta\epsilon} \right]-
\nonumber\\
&&- 2 \c V_I X^I \delta^{i1} \eta_{\alpha [\beta} V_{\gamma]}
+3 \c \epsilon^{1ij}V_I \left[ A^I{}_\alpha J^{(j)}{}_{\beta \gamma}
+{1 \over 3} X^I (\star J^{(j)}){}_{\alpha \beta \gamma} \right],
\end{eqnarray}
which implies
\begin{equation}
 \label{eqn:cclos}
 dJ^{(i)} = 3 \c \epsilon^{1ij} V_I \left( A^I \wedge J^{(j)}+X^I
 \star J^{(j)} \right)
\end{equation}
so $dJ^{(1)}=0$ but $J^{(2)}$ and $J^{(3)}$ are only closed in the
ungauged theory (i.e.\ when $\chi=0$). Equation~(\ref{eqn:cclos})
implies
\begin{equation}
\label{eqn:lied}
\cL_V J^{(i)} = 3 \c \epsilon^{1ij}\left(i_V (V_I A^I) -V_I X^I f \right)
J^{(j)} \,.
\end{equation}
Now consider the effect of a gauge transformation $A^I \rightarrow
A^I+d\Lambda^I$. The Killing spinor equation is invariant provided the
spinor transforms according to
\begin{eqnarray}
\label{eqn:kspmap}
\epsilon^1 &\rightarrow& \cos \left({3 \c V_I \Lambda^I \over 2} \right)
\epsilon^1 - \sin \left({3 \c V_I \Lambda^I \over 2} \right) \epsilon^2
\nonumber\\
\epsilon^2 &\rightarrow& \cos \left({3 \c V_I \Lambda^I \over 2} \right)
\epsilon^2 + \sin \left({3 \c V_I \Lambda^I \over 2} \right) \epsilon^1 \,.
\end{eqnarray}
Under these transformations, $f \rightarrow f$, $V \rightarrow V$ and
$J^{(1)} \rightarrow J^{(1)}$, but $J^{(2)} +i J^{(3)} \rightarrow
e^{-3i \c V_I \Lambda^I} (J^{(2)}+i J^{(3)})$, so $J^{(2,3)}$ are only
gauge-invariant in the ungauged theory. We shall choose to work in a
gauge in which
\begin{equation}
\label{eqn:gaugechoice}
i_V  A^I =f X^I\,.
\end{equation}
In such a gauge we have $\cL_V J^{(i)}=0$.

To make further progress we will examine the dilatino
equation~({\ref{eqn:newdil}}).  Contracting with ${\bar{\epsilon}}^c$
we obtain
\begin{equation}
\label{eqn:liescal}
\cL_V X_I =0
\end{equation}
and
\begin{equation}
\label{eqn:mostuseful}
\left({1 \over 4} Q_{IJ} -{3 \over 8} X_I X_J \right) F^J{}_{\mu \nu}
(J^{(i)} )^{\mu \nu}
=- {3 \c \over 2} \delta^{1i} (X_I V_J X^J -V_I ) f\,.
\end{equation}
Next, contracting~({\ref{eqn:newdil}}) with ${\bar{\epsilon}}^c
\gamma^\sigma$ we find
\begin{equation}
\label{eqn:liegaugeb}
i_V F^J = -d (f X^J )\,,
\end{equation}
which implies that
\begin{equation}
\cL_V F^J=0\,.
\end{equation}
Hence $V$ generates a symmetry of all of the fields.  In the
gauge~({\ref{eqn:gaugechoice}}) we also have
\begin{equation}
\cL_V A^I =0\,.
\end{equation}
Contracting~({\ref{eqn:newdil}}) with ${\bar{\epsilon}}^c
\gamma^\sigma$ we obtain the identity
\begin{equation}
\bigg({1 \over 4} Q_{IJ} -{3 \over 8} X_I X_J \bigg) F^J{}_{\mu \nu}
( \star J^{(i)} )_\sigma{}^{\mu \nu}
=-{3 \over 4} (J^{(i)})_\sigma{}^\mu \nabla_\mu X_I
-{3 \c \over 2} \delta^{i1} (X_I V_J X^J -V_I ) V_\sigma \,.
\end{equation}
Finally, contracting~({\ref{eqn:newdil}}) with  ${\bar{\epsilon}}^c
\gamma^{\sigma \lambda}$ gives
\begin{eqnarray}
\left( {1 \over 4} Q_{IJ} -{3 \over 8} X_I X_J \right)
(F^J{}_{\mu \nu} (\star V)^{\sigma \lambda
\mu \nu}+2 F^{J \lambda \sigma} )
&=&-{3 \over 4}(\nabla^\lambda X_I V^\sigma - \nabla^\sigma X_I
V^\lambda ) +
\nonumber\\
&&+ {3 \c \over 2} (X_I V_J X^J -V_I ) (J^{(1)})^{\sigma \lambda}\qquad
\end{eqnarray}
and
\begin{eqnarray}
\left({1 \over 2} Q_{IJ} -{3 \over 4} X_I X_J \right) \left(F^{J
\sigma}{}_\nu (J^{(i)})^{\nu \lambda}
- F^{J \lambda}{}_\nu (J^{(i)})^{\nu \sigma}\right)
&=& -{3 \over 4} \nabla_\mu X_I (\star J^{(i)})^{\sigma \lambda \mu} +
\\
&& + {3 \c \over 2} \epsilon^{1ij}
\big(X_I V_J X^J -V_I \big) \left(J^{(j)}\right)^{\sigma \lambda}\,.
\nonumber
\end{eqnarray}

\subsection{The timelike case}\label{section2.4}

As in~\cite{gauntlett:02, gauntlett:03}, it is useful to distinguish
two cases depending on whether the scalar $f$ vanishes everywhere or
not. In the former ``null case'', the vector $V$ is globally a null
Killing vector. We shall not consider this case here --- it should be
straightforward to analyze using the methods of~\cite{gauntlett:02,
  gauntlett:03}. In the latter ``timelike case'' there is some open
set ${\cal U}$ in which $f$ is non-vanishing and hence $V$ is a
timelike Killing vector field. There is no loss of generality in
assuming $f>0$ in ${\cal U}$~\cite{gauntlett:02}. We shall analyze the
constraints imposed by supersymmetry in the region ${\cal U}$.

Introduce coordinates $(t,x^m)$ such that $V = \partial/\partial
t$. The metric can then be written locally as
\begin{equation}
 \label{eqn:metric}
 ds^2=f^2(dt+\omega)^2-f^{-1}h_{mn}dx^m dx^n\,.
\end{equation}
The metric $h_{mn}$ can be regarded as the metric on a four
dimensional riemannian manifold, which we shall refer to as the ``base
space" $B$. $\omega$ is a 1-form on $B$. Since $V$ is Killing, $f$,
$\omega$ and $h$ are independent of $t$. We shall reduce the necessary
and sufficient conditions for supersymmetry to a set of equations on
$B$. Let
\begin{equation}
 \label{eqn:e0def}
 e^0 = f (dt+\omega)\,.
\end{equation}
We choose the orientation of $B$ so that $e^0 \wedge \eta_4$ is
positively oriented in five dimensions, where $\eta_4$ is the volume
form of $B$. The two form $d\omega$ can be split into self-dual and
anti-self-dual parts on $B$:
\begin{equation}
\label{eqn:rsp}
f d\omega=G^{+}+G^{-}
\end{equation}
where the factor of $f$ is included for convenience.

Equation~({\ref{eqn:VdotX}}) implies that the $2$-forms $J^{(i)}$ can
be regarded as $2$-forms on the base space and
equation~({\ref{eqn:VstarX}}) implies that they are anti-self-dual:
\begin{equation}
 \star_4 J^{(i)} = - J^{(i)}\,,
\end{equation}
where $\star_4$ denotes the Hodge dual on $B$.
Equation~({\ref{eqn:XcontX}}) can be written
\begin{equation}
 \label{eqn:quat}
 J^{(i)}{}_m{}^p J^{(j)}{}_p{}^n = - \delta^{ij} \delta_m{}^n
 + \epsilon_{ijk} J^{(k)}{}_m{}^n
\end{equation}
where indices $m,n, \ldots$ have been raised with $h^{mn}$, the
inverse of $h_{mn}$. This equation shows that the $J^{(i)}$'s satisfy
the algebra of imaginary unit quaternions, i.e., $B$ admits an almost
hyper-K\"ahler structure, just as in~\cite{gauntlett:02,
  gauntlett:03}.

To proceed, we use~({\ref{eqn:df}}) and~({\ref{eqn:dV}}) to obtain
\begin{eqnarray}
\label{eqn:Fsol}
X_I F^I &=&de^0 -{2 \over 3} G^+ - 2 \c f^{-1} V_I X^I J^{(1)}
\nonumber\\
&=& -f^{-1} e^0 \wedge df +{1 \over 3}G^+ + G^- -2 \c f^{-1}
V_I X^I J^{(1)}\,.
\end{eqnarray}
From~({\ref{eqn:dPhi}}) we find that
\begin{eqnarray}
\label{eqn:nothk}
\nabla_m J^{(1)}_{np} &=& 0
\nonumber\\
\nabla_m J^{(2)}_{np} &=& P_m J^{(3)}_{np}
\nonumber\\
\nabla_m J^{(3)}_{np} &=& -P_m J^{(2)}_{np}\,,
\end{eqnarray}
where $\nabla$ is the Levi-Civita connection on $B$ and we have
defined
\begin{equation}
\label{defbig}
P_m=3 \c V_I (A^I{}_m- f X^I \omega_m )\,.
\end{equation}
From~({\ref{eqn:quat}}) and~({\ref{eqn:nothk}}) we conclude that, in
the gauged theory, the base space is K\"ahler, with K\"ahler form
$J^{(1)}$. In the ungauged theory, it is hyper-K\"ahler with K\"ahler
forms $J^{(i)}$. Again, this is all precisely as in the minimal
theories~\cite{gauntlett:02, gauntlett:03}.

We are primarily interested in the gauged theory so henceforth we
shall assume $\chi \ne 0$.  Proceeding as in~\cite{gauntlett:03}, note
that we can invert~({\ref{eqn:nothk}}) to solve for $P$:
\begin{equation}
\label{eqn:invrta}
P_m ={1 \over 8 } \left( J^{(3) np} \nabla_m J^{(2)}_{np}-
J^{(2) np} \nabla_m J^{(3)}_{np} \right),
\end{equation}
from which it follows that
\begin{equation}
\label{defpee}
dP=\Re\,,
\end{equation}
where $\Re$ is the Ricci-form of the base space $B$ defined by
\begin{equation}
\Re_{mn}={1 \over 2} J^{(1) pq} R_{pq mn}
\end{equation}
and $R_{pqmn}$ denotes the Riemann curvature tensor of $B$.  Hence,
once $B$ has been determined, $P_m$ is determined up to a gradient. An
argument in~\cite{gauntlett:03} shows that the existence of
$J^{(2,3)}$ obeying equations~({\ref{eqn:quat}})
and~({\ref{eqn:nothk}}) is a consequence of $B$ being K\"ahler, and
contains no further information.

Next we examine~({\ref{eqn:mostuseful}}). It is convenient to write
\begin{equation}
\label{eqn:totalident}
F^I = -f^{-1} e^0 \wedge d(fX^I) + \Psi^I + \Theta^I +X^I G^+
\end{equation}
where $\Psi^I$ is an anti-self-dual 2-form on $B$ and $\Theta^I$ is a
self-dual 2-form on $B$. Equation~({\ref{eqn:Fsol}}) implies
\begin{equation}
\label{eqn:gpluscontr}
X_I \Theta^I = -{2 \over 3} G^+
\end{equation}
and
\begin{equation}
X_I \Psi^I = G^- - 2 \c f^{-1} V_I X^I J^{(1)}\,.
\end{equation}
Now ~({\ref{eqn:mostuseful}}) determines $\Psi^I$:
\begin{equation}
\label{eqn:psiexp}
\Psi^I = X^I G^- -9 \c f^{-1} C^{IJK}V_J X_K J^{(1)}\,,
\end{equation}
hence
\begin{equation}
\label{eqn:rewritegaug}
F^I = d (X^I e^0) + \Theta^I  -9 \c f^{-1} C^{IJK}V_J X_K J^{(1)}\,.
\end{equation}
$\Theta^I$ is not constrained by the dilatino equation. Finally,
from~({\ref{defpee}}) together with~({\ref{defbig}}) we have the
following identity
\begin{eqnarray}
3 \c V_I \Theta^I -27 \c^2 f^{-1} C^{IJK}V_I V_J X_K J^{(1)} = \Re\,.
\end{eqnarray}
Contracting this expression with $J^{(1)}$ we obtain
\begin{equation}
\label{eqn:smallfsol}
f =- {108 \c^2 \over R} C^{IJK}V_I V_J X_K\,,
\end{equation}
where $R$ is the Ricci scalar of $B$, and hence
\begin{equation}
\label{eqn:ricform}
\Re -{1 \over 4} R J^{(1)} = 3 \c V_I \Theta^I\,.
\end{equation}
Finally, note that equations~({\ref{eqn:Vproj}})
and~(\ref{eqn:Phiproj}) imply that the spinor obeys the projections
\begin{eqnarray}
\label{eqn:newproj}
\gamma^0 \epsilon^a &=& \epsilon^a\,,
\\
\label{eqn:Jproj}
J^{(1)}_{AB} \Gamma^{AB} \epsilon^a &=& - 4 \epsilon^{ab} \epsilon^b\,,
\end{eqnarray}
where indices $A,B$ refer to a vierbein $e^A$ on the base space, and
$\Gamma^A$ are gamma matrices on the base space given by $\Gamma^A =
\pm i \gamma^A$. These projections are not
independent:~(\ref{eqn:Jproj}) implies~(\ref{eqn:newproj}).

So far we have been discussing constraints on the spacetime geometry
and matter fields that are necessary for the existence of a Killing
spinor. We shall now argue that these constraints are also
sufficient. Assume that we are given a metric of the
form~(\ref{eqn:metric}) for which the base space $B$ is K\"ahler. Let
$J^{(1)}$ denote the K\"ahler form. Assume that $f$ is given in terms
of $X^I$ by equation~(\ref{eqn:smallfsol}) and that the field
strengths are given by equation~(\ref{eqn:rewritegaug}) where
$\Theta^I$ obeys equations~(\ref{eqn:gpluscontr})
and~(\ref{eqn:ricform}). Now consider a spinor $\epsilon^a$ satisfying
the projection~(\ref{eqn:Jproj}). It is straightforward to show this
will automatically satisfy the dilatino
equation~({\ref{eqn:newdil}}). In the basis $(e^0,f^{-1/2} e^A)$, the
gravitino equation~({\ref{eqn:grav}}) reduces to
\begin{equation}
 \partial_t \epsilon^a = 0
\end{equation}
and
\begin{equation}
\label{eqn:kahlerspin}
\nabla_m \eta^a +{1 \over 2}P_m \epsilon^{ab} \eta^b=0\,,
\end{equation}
where
\begin{equation}
\eta^a=f^{-{1 \over 2}} \epsilon^a\,.
\end{equation}
The K\"ahler nature of $B$ guarantees the existence of a solution to
equation~({\ref{eqn:kahlerspin}}) obeying~(\ref{eqn:Jproj}) without
any further algebraic restrictions~\cite{pope:82}.  Therefore the
above conditions on the bosonic fields guarantee the existence of a
Killing spinor, i.e., they are both necessary and sufficient for
supersymmetry. The only projection required is~(\ref{eqn:Jproj}),
which reduces the number of independent components of the spinor from
$8$ to $2$ so we have at least $1/4$ supersymmetry.\footnote{We would
  also expect the general null solution to be $1/4$ supersymmetric, as
  in the minimal theory~\cite{gauntlett:03}. Examples of such
  solutions were given in~\cite{convent:03}. For timelike solutions of
  the ungauged theory, we expect that the only projection that must be
  imposed on a Killing spinor is~(\ref{eqn:newproj}) so the solutions
  will be $1/2$ supersymmetric, as in the minimal
  theory~\cite{gauntlett:02}.}

{\sloppy We have presented necessary and sufficient conditions for
  existence of a Killing spinor. However, we are interested in
  supersymmetric \emph{solutions} so we also need to impose the
  Bianchi identity $dF^I=0$ and Maxwell equations~({\ref{eqn:gauge}}).
  Substituting the field strengths~(\ref{eqn:rewritegaug}) into the
  Bianchi identities $dF^I=0$ gives
\begin{equation}
\label{eqn:bianch}
d  \Theta^I=9 \c C^{IJK} V_J \, d \left(f^{-1} X_K \right)
\wedge J^{(1)}\,.
\end{equation}
Note that
\begin{equation}
\star F^I = -f^{-2} \star_4 d \left(fX^I \right)+e^0 \wedge
\left( \Theta^I+X^I G^+ - \Psi^I \right),
\end{equation}
so the Maxwell equations~({\ref{eqn:gauge}}) reduce to
\begin{eqnarray}
\label{eqn:timegauge}
d \star_4 d \left( f^{-1} X_I \right) &=& -{1 \over 6}C_{IJK} \Theta^J
\wedge \Theta^K +2 \c f^{-1} G^- \wedge J^{(1)} +
\nonumber\\
&& + 6 \c^2 f^{-2}
\left( Q_{IJ}C^{JMN}V_M V_N+V_I X^J V_J \right) \eta_4
\end{eqnarray}
where $\eta_4$ denotes the volume form of $B$.

}

Finally, the integrability conditions for the existence of a Killing
spinor guarantee that the Einstein equation and scalar equations of
motion are satisfied as a consequence of the above equations.

In summary, the general timelike supersymmetric solution is determined
as follows. First pick a K\"ahler 4-manifold $B$. Let $J^{(1)}$ denote
the K\"ahler form and $h_{mn}$ the metric on $B$.
Equation~(\ref{eqn:smallfsol}) determines $f$ in terms of $X^I$. Next
one has to determine $\omega$, $X^I$ and $\Theta^I$ by solving
equations~(\ref{eqn:gpluscontr}),~(\ref{eqn:ricform}),~(\ref{eqn:bianch})
and~(\ref{eqn:timegauge}) on $B$. The metric is then given
by~(\ref{eqn:metric}) and the gauge fields by~(\ref{eqn:rewritegaug}).

\section{Black hole solutions}\label{section3}

\subsection{Derivation of the solutions}\label{section3.1}

Following~\cite{gutowski:04}, we take the following ansatz for the
K\"ahler base space of a supersymmetric black hole solution:
\begin{equation}
 ds_4^2 = d\rho^2 + a^2 \left(
(\sigma_L^{1})^2 + (\sigma_L^{2})^2 \right) + (2 a a')^2 (\sigma_L^{3})^2\,,
\end{equation}
with K\"ahler form
\begin{equation}
 J^{(1)} = - \epsilon d \left[ a^2 \sigma_L^3 \right],
\end{equation}
where $a = a(\rho)$, $\epsilon = \pm 1$, and $\sigma^{i}_L$ are
right-invariant $1$-forms on $\SU(2)$. These can be expressed in terms
of Euler angles $(\theta,\psi,\phi)$ as
\begin{eqnarray}
\label{eqn:sigmadef}
 \sigma_L^{1} &=& \sin \phi d\theta - \cos \phi \sin \theta d\psi
\nonumber\\
 \sigma_L^{2} &=& \cos \phi d\theta + \sin \phi \sin \theta d\psi
\nonumber\\
 \sigma_L^{3} &=& d\phi + \cos \theta d\psi\,,
\end{eqnarray}
where $\SU(2)$ is parametrized by taking $0 \leq \theta \leq \pi$, $0
\leq \phi \leq 4 \pi$ and $0 \leq \psi \leq 2 \pi$.  The
right-invariant 1-forms obey
\begin{equation}
 d\sigma_L^{i} = -\frac{1}{2} \epsilon_{ijk} \sigma_L^{j} \wedge
\sigma_L^{k}\,.
\end{equation}
The surfaces of constant $\rho$ are homogeneous, with a transitively
acting $\U(1)_L \times \SU(2)_R$ isometry group, The $\U(1)_L$
generated by $\partial/\partial \phi$ is manifestly a symmetry and the
$\SU(2)_R$ is a symmetry because $\sigma_L^{i}$ is invariant under the
right action of $\SU(2)$.

We shall assume $a,a'>0$ and introduce an orthonormal basis
\begin{equation}
 e^1 =  d\rho\,,
\qquad
e^2 = a \, \sigma_L^{1}\,,
\qquad
e^3 = a \, \sigma_L^{2}\,,
\qquad
e^4 = 2 a a' \, \sigma_L^{3}
\end{equation}
with volume form $\eta_4 = e^1 \wedge e^2 \wedge e^3 \wedge e^4$.  We
then have
\begin{equation}
 J^{(1)} = -\epsilon( e^1 \wedge e^4 - e^2 \wedge e^3)\,,
\end{equation}
which is obviously anti-self-dual.

As in~\cite{gutowski:04}, we adopt the Ansatz
\begin{equation}
\omega = w (\rho) \sigma_L^3\,,
\end{equation}
which gives
\begin{equation}
G^+ = {f a\over 4a'} \partial_\rho (a^{-2} w) (e^1 \wedge e^4 +e^2 \wedge e^3)
\end{equation}
and
\begin{equation}
G^- = {f \over 4 a^3 a'}\partial_\rho (a^2 w)
(e^1 \wedge e^4 -e^2 \wedge e^3)\,.
\end{equation}
We also take
\begin{equation}
X^I = X^I (\rho)
\end{equation}
and
\begin{equation}
A^I =  X^I e^0  + U^I(\rho) \sigma_L^3\,,
\end{equation}
for some functions $U^I$. $f$ is determined
by~({\ref{eqn:smallfsol}}):
\begin{equation}
f = {54 \c^2 a^2 a' C^{IJK} V_I V_J X_K \over a^2 a''' -a'
+7aa'a''+4 (a')^3}\,.
\end{equation}
Note that
\begin{equation}
\label{eqn:gansatz}
F^I = d(X^I e^0)+ {\partial_\rho U^I \over 2aa'} \, e^1 \wedge
e^4-\frac{U^I}{a^2} \, e^2 \wedge e^3\,.
\end{equation}
Comparing~({\ref{eqn:gansatz}}) with~({\ref{eqn:rewritegaug}}) we find
\begin{equation}
\Theta^I = {a \over 4a'} \partial_\rho (a^{-2} U^I)
(e^1 \wedge e^4 + e^2 \wedge e^3)
\end{equation}
and
\begin{equation}
\label{eqn:uifix}
\partial_\rho (a^2 U^I)= 36 \epsilon \c f^{-1} a^3 a' C^{IJK} V_J X_K\,.
\end{equation}
This equation is sufficient to ensure that the Bianchi
identity~({\ref{eqn:bianch}}) holds
automatically. Equation~({\ref{eqn:gpluscontr}}) gives
\begin{equation}
\label{eqn:calffix}
f^{-1} X_I \partial_\rho (a^{-2} U^I) =-{2 \over 3} \partial_\rho (a^{-2} w)
\end{equation}
and from~({\ref{eqn:ricform}}) we find
\begin{equation}
\label{eqn:kahbaseconstr}
3 \c  V_I U^I = \epsilon (-1 +2aa''+4(a')^2) \,.
\end{equation}
Lastly, we compute the Maxwell equations~({\ref{eqn:timegauge}}); we
obtain
\begin{equation}
\label{eqn:notsonice}
\partial_\rho \left[a^3 a' \partial_\rho (f^{-1} X_I) + \epsilon \c a^2
w V_I +{1 \over 12}C_{IJK}U^J U^K \right]=0\,.
\end{equation}
To find a solution to these equations we make the guess
\begin{equation}
 f^{-1} X_I = \bar{X}_I + \frac{q_I}{4 a^2}\,,
\end{equation}
where $\bar{X}_I$ are the constant values of the scalars in the
maximally supersymmetric $AdS_5$ solution and $q_I$ are
constants. Since
\begin{equation}
 C^{IJK}X_I X_J X_K = \frac{2}{9}
\end{equation}
we must have
\begin{equation}
 f^{-3} = \frac{9}{2} C^{IJK} \left( \bar{X}_I + \frac{q_I}{4 a^2}
\right) \left( \bar{X}_J + \frac{q_J}{4 a^2} \right)
\left( \bar{X}_K + \frac{q_K}{4 a^2}\right),
\end{equation}
which implies
\begin{equation}
 f = \left( 1 + \frac{\alpha_1}{4 a^2} + \frac{\alpha_2}{16 a^4} +
\frac{\alpha_3}{64 a^6} \right)^{-1/3},
\end{equation}
where we have defined the constants
\begin{equation}
\label{eqn:alphadef}
 \alpha_1 = \frac{27}{2} C^{IJK} \bar{X}_I \bar{X}_J q_K\,,
\qquad
\alpha_2 = \frac{27}{2} C^{IJK} \bar{X}_I q_J q_K\,,
\qquad
\alpha_3 = \frac{9}{2} C^{IJK} q_I q_J q_K\,.
\end{equation}
Equation~(\ref{eqn:uifix}) then determines
\begin{equation}
 U^I = \frac{9 \epsilon}{\ell} C^{IJK} \bar{X}_J \left( a^2 \bar{X}_K
+ \frac{q_K}{2} \right).
\end{equation}
A possible term of the form (constant of integration) times $a^{-2}$
has been set to zero because such a term is not present in the
supersymmetric black hole solutions of the minimal theory.  Now we can
determine $w$ from~(\ref{eqn:calffix}):
\begin{equation}
 w = \frac{\epsilon}{\ell} \left( w_0 a^2 - \frac{\alpha_1}{2} -
\frac{\alpha_2}{16 a^2} \right),
\end{equation}
where $w_0$ is a constant of integration.
Equation~(\ref{eqn:kahbaseconstr}) can be integrated to give
\begin{equation}
 (a')^2 = \frac{a^2}{\ell^2} + \frac{1}{4} \left( 1 +
\frac{\alpha_1}{\ell^2} \right) + \frac{\kappa}{a^4}\,,
\end{equation}
where $\kappa$ is a constant of integration. Comparison with the
minimal theory shows that we must take $\kappa=0$. We can then
integrate to obtain
\begin{equation}
 a = \frac{\ell}{2} \sqrt{ 1 + \frac{\alpha_1}{\ell^2} }
\sinh\left(\frac{\rho}{\ell}\right).
\end{equation}
This determines the geometry of the base space: it is the same
singular deformation of the Bergmann manifold that appears in the
minimal theory~\cite{gutowski:04}.  Finally, we require
equation~(\ref{eqn:notsonice}) to be satisfied. Upon using
equation~(\ref{eqn:jordan}), this reduces to
\begin{equation}
 w_0 = -2\,.
\end{equation}
Hence
\begin{equation}
 w = - \frac{\epsilon}{\ell} \left( 2 a^2 + \frac{\alpha_1}{2} +
\frac{\alpha_2}{16 a^2} \right).
\end{equation}

\subsection{Properties of the solution}\label{section3.2}
\label{subsec:properties}

It is convenient to work with a new radial coordinate $R$ defined
by\footnote{Note that this differs from the coordinate $R$ used
  in~\cite{gutowski:04} by a factor of $f$.}
\begin{equation}
 a = \frac{R}{2}.
\end{equation}
The metric is
\begin{equation}
 ds^2 = f^2 dt^2 + 2 f^2 w \, dt \, \sigma_L^3 - f^{-1} g^{-1} dR^2 -
\frac{R^2}{4} \left[ f^{-1} ( (\sigma_L^1)^2 + (\sigma_L^2)^2
) + f^2 h (\sigma_L^3)^2  \right],
\end{equation}
where $\sigma_L^i$ was defined in equation~(\ref{eqn:sigmadef}) and
\begin{eqnarray}
 f &=& \left( 1 + \frac{\alpha_1}{R^2} + \frac{\alpha_2}{R^4} +
\frac{\alpha_3}{R^6} \right)^{-1/3},
\\
 w &=& - \frac{\epsilon R^2}{2 \ell} \left( 1 + \frac{\alpha_1}{R^2}
+ \frac{\alpha_2}{2 R^4} \right),
\\
 g &=& 1 + \frac{\alpha_1}{\ell^2} + \frac{R^2}{\ell^2}\,,
\\
 h &\equiv & f^{-3} g - \frac{4}{R^2} w^2 =  1 +
\frac{\alpha_1}{R^2} + \frac{1}{R^4} \left( \alpha_2 + \frac{
\alpha_3}{\ell^2} \right) + \frac{1}{R^6} \left[ \alpha_3 \left( 1 +
\frac{ \alpha_1}{\ell^2} \right) - \frac{\alpha_2^2}{4 \ell^2} \right],\quad
\end{eqnarray}
where $\epsilon= \pm 1$ determines the sense of rotation.  The scalars
are
\begin{equation}
 X_I = f \left( \bar{X}_I + \frac{q_I}{R^2} \right).
\end{equation}
The gauge fields are
\begin{equation}
 A^I = f X^I dt + \left(U^I + f w X^I \right) \sigma_L^3\,,
\end{equation}
where
\begin{equation}
 U^I = \frac{9\epsilon}{4 \ell} C^{IJK} \bar{X}_J \left( \bar{X}_K R^2
+ 2 q_K  \right).
\end{equation}
The solution is parametrized by the constants $q_I$, which determine
$\alpha_i$ via equation~(\ref{eqn:alphadef}). Recall that the
constants $\bar{X}_I$ are the values of the scalars in the $AdS_5$
vacuum solution, which has radius $\ell$. The solution has isometry
group $R \times \U(1)_L \times \SU(2)_R$ where $R$ is generated by the
supersymmetric Killing vector field $V=\partial/\partial t$, $\U(1)_L$
by $\partial/\partial \phi$ and the $\SU(2)_R$ arises because
$\sigma_L^i$ is invariant under the right action of $\SU(2)$.  The
supersymmetric black hole solutions of the minimal theory
have~\cite{gutowski:04} $\alpha_1 = 3 R_0^2$, $\alpha_2 = 3 R_0^4$ and
$\alpha_3 = R_0^6$.

To see that the solution is asymptotically AdS, let
\begin{equation}
 \phi' = \phi + \frac{2 \epsilon t }{\ell}\,.
\end{equation}
We then have
\begin{equation}
\label{eqn:asympads}
 ds^2 = f^{-1} g h^{-1} dt^2  - f^{-1}
g^{-1} dR^2 - \frac{R^2}{4} \left[ f^{-1} \left( (\sigma_L^{1'})^2 +
(\sigma_L^{2'})^2 \right) + f^2 h (\sigma_L^{3'} -
\Omega dt )^2 \right],
\end{equation}
where $\sigma_L^{i'}$ is defined in the same way as $\sigma_L^i$
(equation~(\ref{eqn:sigmadef})) but with $\phi$ replaced by $\phi'$
and
\begin{equation}
 \Omega = \frac{2\epsilon}{\ell h} \left\{ \frac{1}{R^4} \left(
\frac{\alpha_2}{2} + \frac{ \alpha_3}{\ell^2} \right) + \frac{1}{R^6}
\left[ \alpha_3  \left( 1 +
\frac{ \alpha_1}{\ell^2} \right) - \frac{\alpha_2^2}{4 \ell^2}
\right] \right\}.
\end{equation}
Now as $R\rightarrow \infty$, we have $f,h \rightarrow 1$, $\Omega
\rightarrow 0$ and $g \propto R^2/\ell^2$ so the solution is
manifestly asymptotic to $AdS_5$. In these coordinates, the
supersymmetric Killing vector field is $V = \partial/\partial t + (2
\epsilon / \ell) \partial /\partial \phi'$. The gauge fields have the
asymptotic behaviour
\begin{eqnarray}
 A^I &=& \bigg\{ \bar{X}^I - \frac{1}{R^2} \left[ \left( \alpha_1 +
\frac{\alpha_2}{2\ell^2} \right) \bar{X}^I -
 9 C^{IJK} \bar{X}_J q_K -
\frac{9}{2\ell^2} C^{IJK} q_J q_K \right] +
\nonumber\\
&&\hphantom{\bigg\{}\!
+ {\cal O}\left( \frac{1}{R^4} \right) \bigg\} dt
+ {\cal O} \left( \frac{1}{R^2} \right) \sigma_L^{3'}\,.
\end{eqnarray}
Note that if $\alpha_2 = \alpha_3 = 0$ then the solution reduces to
one of the static, spherically symmetric, nakedly singular, solutions
investigated in~\cite{behrndt:98}.

To investigate which solutions have regular horizons, we shall attempt
to introduce gaussian null coordinates as follows:
\begin{equation}
dt = du -f g^{-1} h \, dr\,,
\qquad
d\phi = d\phi'' - \frac{4}{R^2} f g^{-1} w  \, dr\,,
\qquad
dR = f \sqrt{h} \, dr\,.
\end{equation}
The line element becomes
\begin{equation}
 ds^2 = f^2 du^2 - 2 du dr + 2 f^2 w \, du \, \sigma_L^{3''} -
 \frac{R^2}{4} \left[ f^{-1} \left( (\sigma_L^{1''})^2 +
 (\sigma_L^{2''})^2 \right) + f^2 h (\sigma_L^{3''})^2 \right],
\end{equation}
where $\sigma_L^{i''}$ is defined in the same way as $\sigma_L^i$
(equation~(\ref{eqn:sigmadef})) but with $\phi$ replaced by $\phi''$.
The supersymmetric Killing vector field is $V = \partial/\partial u$.
In order for there to be a regular horizon at $R=0$ we need $R^2
f^{-1}$ to approach a positive constant as $R \rightarrow 0$. This
requires
\begin{equation}
 \alpha_3 > 0\,.
\end{equation}
We also need $R^2 f^2 h$ to approach a positive constant, which
requires
\begin{equation}
\label{eqn:bhconstr}
 \alpha_3 \left( 1 +
\frac{ \alpha_1}{\ell^2} \right) - \frac{\alpha_2^2}{4 \ell^2} > 0\,.
\end{equation}
We then find that $r \propto R^2$ as $R \rightarrow 0$ and that $f$
and $f^2 w$ are ${\cal O}(r)$ as $r \rightarrow 0$, which guarantees a
regular horizon. Hence, subject to the above restrictions, our
solution has a regular horizon at $R=0$. In order to avoid problems in
$R>0$ we must also demand that $f$, $g$ and $h$ be positive for $R>0$,
which imposes further restrictions on $\alpha_i$.

Spatial cross-sections of the event horizon have the geometry of a
squashed $S^3$ with area
\begin{equation}
 {\cal A} = 2 \pi^2 \sqrt{ \alpha_3 \left( 1 + \frac{\alpha_1}{\ell^2}
\right) - \frac{\alpha_2^2}{4\ell^2}}\,.
\end{equation}
The angular velocity of the event horizon with respect to the
stationary frame at infinity can be calculated as
in~\cite{gutowski:04}, giving
\begin{equation}
 \Omega_H = \frac{2 \epsilon}{\ell}\,.
\end{equation}
We note that if we take the limit $\ell \rightarrow \infty$ with $q_I$
held fixed then our solutions reduce to \emph{static} supersymmetric
black hole solutions of the ungauged supergravity theory, which were
first obtained in~\cite{sabra:98}.

We can calculate the mass and angular momentum of our solutions using
the definitions of Ashtekar and Das (AD)~\cite{ashtekar:00}. The AD
mass is associated with the symmetry of the conformal boundary
generated by $\partial /\partial t$ in the
coordinates~(\ref{eqn:asympads}):
\begin{equation}
 M = \frac{\pi}{4G} \left( \alpha_1 + \frac{3 \alpha_2}{2
\ell^2} + \frac{2 \alpha_3}{\ell^4} \right),
\end{equation}
and the AD angular momentum is associated with $-\partial/\partial
\phi'$:
\begin{equation}
 J = \frac{\epsilon \pi}{8 G \ell} \left( \alpha_2 + \frac{2
\alpha_3}{\ell^2} \right).
\end{equation}
Just as in the minimal theory, this really corresponds to equal
angular momenta $J_1 = J_2 = J$ in two orthogonal 2-planes.

We have used the AD approach to define mass and angular
momentum. However, it has been argued~\cite{deharo:01, skenderis:01}
that the conserved quantities of the AD approach do not correctly
reproduce the (anomalous) transformation law of the CFT
energy-momentum tensor~\cite{henningson:98}, so the AD mass $M$ should
not be interpreted as dual to CFT energy. Instead one should use an
alternative approach based on the ``holographic stress tensor"
(HST)~\cite{balasubramanian:99, kraus:99}, which does transform
correctly.

If $M$ cannot be interpreted as dual to CFT energy then what is its
CFT interpretation? To answer this, we first note that the AD
definitions only apply to spacetimes that are asymptotically AdS
whereas the HST approach applies more generally to spacetimes that are
merely asymptotically locally AdS with a well-defined conformal
boundary.  Therefore, when $M$ is defined, the dual CFT must live on
$R \times S^3$ whereas the HST energy $E$ can be defined for many
different CFT background geometries.  Now $R \times S^3$ is precisely
the background for which the CFT state/operator correspondence
applies~\cite{witten:98}.  Under this correspondence, the energy of a
state is equal to the dimension of the corresponding operator plus an
anomalous term that can be interpreted as the Casimir energy on $R
\times S^3$. This suggests that we should identify the AD mass of a
bulk solution with the dimension of the corresponding local CFT
operator(s):
\begin{equation}
 \Delta = M \ell\,.
\end{equation}
Of course, this should only be regarded as the leading term in a large
$N$ expansion. More precisely, we mean that $\Delta/N^2$ and $M
\ell/N^2$ tend to the same limit as $N \rightarrow \infty$.

In the few examples for which $M$ and $E$ have both been calculated,
it has been found that they differ precisely by the Casimir energy of
the CFT on $R \times S^3$, which is evidence in favour of the above
interpretation.\footnote{It was claimed in~\cite{das:00} that the
difference is more complicated for the rotating black hole solutions
of~\cite{hawking:99}. However, the HST results of~\cite{das:00}
(following the earlier~\cite{awad:00}) correspond to the stress tensor
of a CFT in $R \times S^3$ with a non-product metric~\cite{awad:00},
for which one would expect the Casimir energy to be more complicated
anyway.}  It would be interesting to see whether this could be proved
more generally. It would also be interesting to calculate the HST for
our solutions. This was done for solutions of the minimal theory
in~\cite{gutowski:04}, where it was found that $M$ and $E$ differ by
the Casimir energy and that the two approaches give the same value for
$J$. In the theory under consideration here, the counterterms required
for calculating the HST do not appear to have been derived yet.

We shall define conserved electric charges by:
\begin{equation}
 Q_I = \frac{1}{8\pi G} \int Q_{IJ} \star F^J\,,
\end{equation}
where the integral is taken over a spatial three sphere at
infinity. Calculating this on a surface of constant $t$ in the
asymptotically AdS coordinates gives
\begin{equation}
 Q_I = \frac{\pi}{G} \left( \frac{3}{4} q_I - \frac{3 \alpha_2
 }{8 \ell^2} \bar{X}_I + \frac{9}{8\ell^2}C_{IJK} \bar{X}^J C^{KLM} q_L
 q_M \right).
\end{equation}
This implies
\begin{equation}
 \bar{X}^I Q_I = \frac{\pi}{4 G} \left( \alpha_1 +
\frac{\alpha_2}{2\ell^2} \right),
\end{equation}
so we have the BPS equality
\begin{equation}
 M - \frac{2}{\ell} |J| = |\bar{X}^I Q_I |\,.
\end{equation}
It would be interesting to look for non-extremal generalizations of
these solutions. In general, such solutions will carry two independent
angular momenta, which will make them rather complicated
(see~\cite{hawking:99} for uncharged solutions). However, the
solutions should simplify when the angular momenta are equal, with the
isometry group being enhanced from $R \times \U(1)^2$ to $R \times
\U(1)_L \times \SU(2)_R$ (as for the supersymmetric solutions). If the
metric is written using right-invariant forms on $\SU(2)$ then the
metric components will all be functions of a single radial coordinate,
so finding these solutions should not be difficult.

\subsection{Solutions of the $\U(1)^3$ theory}\label{section3.3}

We are primarily interested in solutions that can be oxidized to yield
asymptotically $AdS_5 \times S^5$ solutions of type-IIB
supergravity. We therefore consider the theory with $\U(1)^3$ gauge
group obtained by taking indices $I,J,K$ to run from $1$ to $3$ and
with $C_{IJK} = 1$ if $(IJK)$ is a permutation of $(123)$ and $C_{IJK}
= 0$ otherwise. The constraint on the scalars is then
\begin{equation}
 X^1 X^2 X^3 = 1
\end{equation}
and we have
\begin{equation}
 Q_{IJ} = \frac{9}{2} {\rm diag} \left( (X_1)^2,(X_2)^2,(X_3)^2 \right).
\end{equation}
We also take
\begin{equation}
 V_I = \frac{\xi}{3}\,.
\end{equation}
Solutions of this theory can be oxidized to solutions of type-IIB
supergravity as described in~\cite{cvetic:99}.\footnote{Our gauge
  fields have the same normalization as those
  of~\cite{cvetic:99}. Denote the scalars of~\cite{cvetic:99} by
  $\tilde{X}_I$. They are related to our scalars by $\tilde{X}_I =
  1/(3 X_I)$. The coupling of~\cite{cvetic:99} is related to our
  constants by $g = \chi \xi = 1/\ell$.}  In this theory,
\begin{equation}
 \bar{X}_I = \frac{1}{3}\,,
\qquad
\bar{X}^I = 1\,.
\end{equation}
It is convenient to define rescaled parameters for the black hole
solutions:
\begin{equation}
 q_I = \frac{\mu_I}{3}
\end{equation}
so that
\begin{equation}
\alpha_1 = \mu_1 + \mu_2 + \mu_3\,,
\qquad
\alpha_2 = \mu_1 \mu_2 + \mu_2 \mu_3 + \mu_3 \mu_1\,,
\qquad
\alpha_3 = \mu_1 \mu_2 \mu_3\,.
\end{equation}
We then have
\begin{equation}
 f = \left( H_1 H_2 H_3 \right)^{-1/3},
\end{equation}
where
\begin{equation}
 H_I = 1 + \frac{\mu_I}{R^2}\,.
\end{equation}
The scalars are given by
\begin{equation}
 X_I = \frac{1}{3} H_I \left( H_1 H_2 H_3 \right)^{-1/3}.
\end{equation}
We need $\mu_I > 0$ to guarantee $\alpha_3 >0$ and $f>0$ for $R>0$.
The only remaining restriction on $\mu_I$ for the solution to describe
a black hole is equation~(\ref{eqn:bhconstr}), which can be written
\begin{equation}
\label{eqn:newconstr}
 4 \mu_1 \mu_2 \mu_3 \left( \mu_1 + \mu_2 + \mu_3 + \ell^2 \right) >
\left( \mu_1 \mu_2 + \mu_2 \mu_3 + \mu_3 \mu_1 \right)^2.
\end{equation}
This constraint is non-trivial: e.g.\ it is not satisfied if we take $
\mu_1 = \mu_2 \gg \mu_3$. The mass, angular momentum and charges can
be obtained from the expressions above. The charges simplify to
\begin{eqnarray}
 Q_1 &=& \frac{\pi}{4G} \left[ \mu_1 + \frac{1}{2\ell^2} \left( \mu_1
\mu_2 + \mu_1 \mu_3 - \mu_2 \mu_3 \right) \right]
\nonumber\\
 Q_2 &=& \frac{\pi}{4G} \left[ \mu_2 + \frac{1}{2\ell^2} \left( \mu_2
\mu_3 + \mu_2 \mu_1 - \mu_3 \mu_1 \right) \right]
\nonumber\\
 Q_3 &=& \frac{\pi}{4G} \left[ \mu_3 + \frac{1}{2\ell^2} \left( \mu_3
\mu_1 + \mu_3 \mu_2 - \mu_1 \mu_2 \right) \right].
\end{eqnarray}
It is possible to set one, but not two, of the charges to zero
consistently with the constraint~(\ref{eqn:newconstr}).

\acknowledgments

This research was supported in part by the National Science Foundation
under Grant No. PHY99-07949. J. B. G. was supported by EPSRC. HSR
thanks Radu Roiban for useful discussions.

\end{document}